\documentclass{article}
\usepackage{graphicx}
\usepackage{epsfig}
\input amssym.def
\input amssym.tex

\def\lb{\label}
\def\ben{\begin{eqnarray}}
\def\een{\end{eqnarray}}
\def\be{\begin{eqnarray}}
\def\ee{\end{eqnarray}}

\def\bea{\begin{eqnarray}}
\def\eea{\end{eqnarray}}

\def\Tr{{\rm Tr}\, }
\def\appendix{{\newpage\section*{Appendix}}\let\appendix\section%
        {\setcounter{section}{0}
        \gdef\thesection{\Alph{section}}}\section}
\newcount\hour \newcount\minute
\hour=\time  \divide \hour by 60 \minute=\time \loop \ifnum
\minute > 59 \advance \minute by -60 \repeat
\def\nowtwelve{\ifnum \hour<13 \number\hour:
                      \ifnum \minute<10 0\fi
                      \number\minute
                      \ifnum \hour<12 \ A.M.\else \ P.M.\fi
         \else \advance \hour by -12 \number\hour:
                      \ifnum \minute<10 0\fi
                      \number\minute \ P.M.\fi}
\def\nowtwentyfour{\ifnum \hour<10 0\fi
                \number\hour:
                \ifnum \minute<10 0\fi
                \number\minute}

\begin{document}

\hfuzz=100pt
\title{The Geometry of Large Causal Diamonds and the No Hair Property of Asymptotically de-Sitter Spacetimes}
\author{G. W. Gibbons
\\
D.A.M.T.P.,
\\Cambridge University,
\\Wilberforce Road, Cambridge CB3 0WA, U.K.and
\\The Galileo Galilei Institute for Theoretical Physics
\\Arcetri, Firenze, Italy
\\
\\ S. N. Solodukhin
\\Arnold-Sommerfeld-Center for Theoretical Physics,
\\Department f\"{u}r Physik,
\\Ludwig-Maximilians Universit\"{a}t,
\\Theresienstrasse 37, D-80333, M\"{u}nchen, Germany}

\maketitle


\begin{abstract}
In a previous paper we obtained formulae for the volume of a
causal diamond or Alexandrov open set  $I^+(p) \cap I^-(q)$ whose
duration $\tau(p,q) $ is short compared with the curvature scale.
In the present paper we obtain asymptotic formulae valid  when the
point $q$ recedes to the future boundary ${\cal I}^+$ of an
asymptotically de-Sitter spacetime. The volume (at fixed $\tau$)
remains finite in this limit and is  given by the universal
formula $V(\tau) = {4\over 3}\pi (2\ln \cosh{\tau\over
2}-\tanh^2{\tau\over 2})$ plus  corrections (given by a series in
$e^{-t_q}$ ) which  begin at order $e^{-4t_q}$. The coefficents of
the corrections depend on the geometry of ${\cal I}^+$. This
behaviour is shown to be consistent  with the no-hair property  of
cosmological event horizons and with calculations of de-Sitter
quasinormal modes in the literature.

\end{abstract}

\section{Introduction}
In a recent paper \cite{GibbonsSolodukhin} we embarked on a
quantitative study of causal diamonds, or Alexandrov open sets,
which are beginning to play an increasingly important role in
quantum gravity, for example in the approach via casual sets
\cite{Sorkin},  in  discussions of \lq holography\rq , and also of
the probability of various observations in eternal inflation
models (see \cite{Bousso} for a recent example and references to
earlier work). The calculations in \cite{GibbonsSolodukhin} were
concerned with small causal diamonds,  that is causal diamonds
$I^+(p) \cap I^-(q)$ whose duration $\tau(p,q)$ \footnote{For
relevant definitions and notation the reader is directed to
\cite{GibbonsSolodukhin}.}  is small compared with the ambient
curvature scale. The  present  paper was  motivated by
inflationary cosmology and the observations showing that the scale
factor $a(t)$ of our present universe is accelerating. Indeed, it
is given to a good approximation by assuming that the spatial
geometry is flat and setting  the \ben {\rm jerk} \equiv  { a^2
\over {\dot a}^3 } { d^3 a \over d t ^3 }  =1 \,
 \een
so that \ben a(\tau)=  \sinh ^{ 2 \over 3} \bigl (  \sqrt{3
\Lambda \over 4} t \bigr )\,,\label{scale}\een where $\Lambda$ is
the cosmological constant. The jerk is a dimensionless measure of
the rate  of change of acceleration. It is easily seen to be unity
if and only if we have   $k=0$ model with a cosmological constant
and pressure free matter \cite{Chiba, Alam, Sahni, Blandford}. The
physical reason why the jerk of the observed  universe is unity is
unclear. Equation (\ref{scale}) then solves Einstein's equations
with a cosmological term coupled to a  pressure free fluid.

The questions  we are interested in concern the observations made
by a  hypothetical observer moving along a timelike world line
$\gamma$, in  metric  which is not exactly but only asymptotically
de-Sitter, in the limit that his/her  own proper time
$t_q\rightarrow \infty$. In particular we shall study the volume
$V(\tau, t_q)$ of the causal diamond $I^+(p) \cap I^-(q)$ where
$p$ and $q$ lie on $\gamma$ in the limit when both $t_p, \ t_q
\rightarrow \infty$ while $\tau=t_q-t_p$ is kept fixed. Thus both
points $p$ and $q$ tend to future spacelike infinity ${\cal I}^+$
while the duration of the diamond $\tau$ is kept fixed. The entire
diamond is in the asymptotic region and the volume of the diamond
depends on the asymptotic geometry which we wish to explore.
 The volume (at fixed $\tau$)  remains finite in this limit
and is  given by the universal formula $V(\tau) = {4\over 3}\pi
(2\ln \cosh{\tau\over 2}-\tanh^2{\tau\over 2})$ plus corrections
(which are given by a  series in $e^{-t_q}$ ) which begin at order
$e^{-4t_q}$. This behaviour will be shown to be consistent  with
the no-hair property  of cosmological event horizons and with
calculations of  de-Sitter  quasinormal modes in the literature.

Before describing our calculations,we shall give a brief review of
the geometry of asymptotically de-Sitter spacetimes.

\section{Geometry of asymptotically de-Sitter spacetimes}

>From now on we adopt units in which $\Lambda=3$ and hence $H=1$.
The metric on de-Sitter space  may be cast in Friedmann-Lemaitre
form in three different ways \bea
(i)& & \qquad k=+1\,,\qquad a(t) = \cosh t\,,\\
(ii)& & \qquad k=\,\, \,0\,,\qquad \, a(t)= \exp t \,,\\
(iii)& & \qquad k=-1\, ,\qquad a(t) = \sinh t \,. \eea Of these
only the first is global, that is covers the full geodesically
complete spacetime. Another local chart, valid only  inside an
observer dependent
 cosmological horizon,   is the locally static form
\ben ds ^2 =-(1-r^2) dt^2 + {dr ^2 \over 1-r^2 } + r^2 \bigl (d
\theta ^2 + \sin ^2 \theta d \phi ^2 \bigr )\,. \label{static}
\een It was conjectured in \cite{GibbonsHawking}, before the
theory of inflation,  that perturbations of de-Sitter should
settle down inside the the cosmological event horizon ($r<1$) to
the exact static form. How this \lq No Hair \rq mechanism works in
practice was later elucidated in the context of inflation in
\cite{BoucherGibbons} who pointed out that while scalar and
gravitational perturbations of de-Sitter spacetime described
using any of the Friedmann-Lemaitre coordinates do not decay, but
rather freeze in to constant values at late times, restricted to
interior the event horizon of any given inertial observer, the
perturbations decay exponentially. One way of understanding  this
is to note \cite{Starobinsky} that the general asymptotic form of
the metric at late times expressed in quasi-Friedmann-Lemaitre,
geodesic or Gaussian  coordinates takes the form \ben ds ^2 =-dt
^2 +  e^{2t} g_{ij} (x) dx^i dx ^j + \dots \label{late} \een where
$g_{ij}$ is an {\sl arbitrary} three-metric.

Thus {\sl globally} the metric does not settle down to the
de-Sitter form. However {\it locally} that is within the event
horizon of any given observer it does. That is because as time
goes on, such an observer can access an exponentially smaller and
smaller  proportion  of the spatial hypersurface $ \Sigma: t={\rm
constant}$. Now provided that the $\Sigma$ is smooth, no matter
what metric it is given, any local patch when examined with
sufficient magnification will appear flat. Exact solutions of the
Einstein  equations describing this process are rather rare, but
there are some: the Biaxial  Taub-NUT metrics, and that exhibits
this mechanism rather clearly \cite{GibbonsRuback}. For a recent
astrophyisical perspective on the eschatology of an asymptotically
de-Sitter universes see \cite{KraussSherrer}.

In a later, and  completely independent, development Fefferman and
Graham \cite{FeffermanGraham} examined  asymptotically hyperbolic
Riemannian (i.e positive definite) Einstein metrics with negative
scalar curvature near their conformal boundary. It is clear that
the asymptotic expansions they obtained are identical in structure
to those discussed by Starobinsky earlier \cite{Starobinsky} for a
Lorentzian Einstein metrics with positive  scalar curvature near
its spacelike conformal boundary. They are also identical in
structure to the  asymptotical anti-de-Sitter metrics near their
timelike boundary \cite{deHaroSolodukhinSkenderis}. In what
follows we shall make use of these expansions. For more work on
the de-Sitter case see \cite{Rendall}.

We consider a ($d+1)$-dimensional space-time which solves the
Einstein equations \be R_{\mu\nu}-{1\over 2}G_{\mu\nu}R=\Lambda
G_{\mu\nu} \lb{1} \ee with negative cosmological constant
$\Lambda={d(d-1)\over 2l^2}$, $l$ is the de-Sitter radius. We look
at the solution to these equations close to the spacelike infinity
${\cal I}^+$ in the form \be ds^2=-l^2{d\rho^2\over
2\rho^2}+{1\over \rho}g_{ij}(x,\rho)dx^idx^j \lb{2} \ee where
$\rho$ is a timelike coordinate such that $\rho=0$ at ${\cal
I}^+$. Coordinates $x^i, \ i=1,..,d$ are the coordinates on the
spacelike surface ${\cal I}^+$. Inserting this metric into the
Einstein equations one obtains a system of equations \bea \rho
\,[2 g^{\prime\prime} - 2 g^\prime g^{-1} g^\prime + \Tr\, (g^{-1}
g^\prime)\, g^\prime] + l^2{\rm Ric} (g) - (d - 2)\, g^\prime -
\Tr \,(g^{-1} g^\prime)\, g =  0 \cr \nabla_i\, \Tr \,(g^{-1}
g^\prime) - \nabla^j g_{ij}^\prime   =  0 \cr \Tr \,(g^{-1}
g^{\prime\prime}) - \frac{1}{2} \Tr \,(g^{-1} g^\prime g^{-1}
g^\prime) =  0 , \label{3} \eea where differentiation with respect
to $\rho$ is denoted with a prime, $\nabla_i$ is the covariant
derivative constructed from the metric $g$, and ${\rm Ric} (g)$ is
the Ricci tensor of $g$.

Notice that we could have considered the Einstein equations with
negative cosmological constant $\Lambda={-d(d-1)/2l^2}$. The
analytic continuation between two cases is a simple replacement
$l^2\rightarrow -l^2$ both in the metric (\ref{2}) and in
equations (\ref{3}). The analytic continuation between two  spacetimes was
considered in detail in the appendix of \cite{Skenderis}.
Coordinate $\rho$ then becomes a radial
coordinate, $\rho=0$ is the timelike infinity of the
asymptotically anti-de-Sitter space-time. The solution of the
equations (\ref{3}) in this case is well known in the form of the
asymptotic expansion \be
g_{ij}(x,\rho)=g^{(0)}_{ij}(x)+g^{(2)}_{ij}(x)\rho+..+g^{(d)}_{ij}(x)\rho^{d/2}+h^{(d)}_{ij}(x)\rho^{d/2}\ln\rho+..
\lb{4} \ee where $g^{(0)}_{ij}(x)$ is the metric on the timelike
boundary of the anti-de-Sitter space-time. The coefficients
$g^{(k)}_{ij}(x)$, $k<d$ and $h^{(d)}_{ij}(x)$ are uniquely
determined by the metric $g^{(0)}_{ij}(x)$ while for
$g^{(d)}_{ij}$ only the trace and the covariant divergence are
determined by $g^{(0)}_{ij}(x)$. $g^{(d)}_{ij}$ thus encodes the
stress energy tensor of the boundary dual theory. Coefficient
$h^{(d)}_{ij}(x)$ is non-vanishing only if $d$ is even, it has
some interesting conformal properties and mathematicians call it
the obstruction tensor.

In the asymptotically de-Sitter case one can use same expansion
(\ref{4}) taking into account that $\rho$ is now a time-like
coordinate and metric $g^{(0)}_{ij}(x)$ is now the metric on the
spacelike future infinity ${\cal I}^+$. Moreover, all expressions
for the coefficients $g^{(k)}_{ij}(x)$ and $h^{(d)}_{ij}(x)$ as
determined by $g^{(0)}_{ij}(x)$ take exactly same form as in the
asymptotically anti-de-Sitter case provided the substitution
$l^2\rightarrow -l^2$ is applied.  In particular we find for the
first few coefficients\footnote{Notice that our curvature
notations differ by a sign from those used in
\cite{Henningson:1998gx} and \cite{deHaroSolodukhinSkenderis}.}
\cite{Henningson:1998gx}, \cite{deHaroSolodukhinSkenderis}, \be
&&g^{(2)}_{ij}(x)={l^2\over
(d-2)}(R_{ij}-{1\over 2(d-1)}Rg^{(0)}_{ij}) \, , \nonumber \\
&&g^{(4)}_{ij}
 =  \frac{l^4}{(d - 4)} \left(  \frac{1}{8 (d - 1)} \nabla_i \nabla_j R
 -\frac{1}{4 (d - 2)} \nabla_k \nabla^k R_{ij} \right . \cr & & +
\frac{1}{8 (d - 1) (d - 2)} \nabla_k \nabla^k R g^{(0)}_{ij} -
\frac{1}{2 (d - 2)} R^{kl} R_{ikjl} \cr & & + \frac{d - 4}{2 (d -
2)^2} R_i{}^k R_{kj} + \frac{1}{(d - 1)(d - 2)^2} R R_{ij} \cr & &
\left. + \frac{1}{4 (d - 2)^2} R^{kl} R_{kl} g^{(0)}_{ij} -
\frac{3 d}{16 (d - 1)^2 (d - 2)^2} R^2 g^{(0)}_{ij} \right) \lb{5}
\ee in the asymptotically de-Sitter case. The expressions for
$g^{(k)}_{ij}$ are singular when $k=d$. In this case
$g^{d)}_{ij}(x)$ is not uniquely determined by metric
$g^{(0)}_{ij}$. The Einstein equations impose certain constraints
on the trace and covariant divergence of coefficient
$g^{(d)}_{ij}(x)$.

\section{Volume  of the causal diamond}

In the rest of the paper we will be interested in a
four-dimensional asymptotically de-Sitter space-time so that
$d=3$. Since $d$ is odd no obstruction tensor appears in the
expansion (\ref{4}). From now on we use units in which $l=1$.

\bigskip

\noindent{\it Asymptotic metric.} In  addition to $\rho$,  two
other timelike coordinates can be used. The coordinate $t$ is
defined by relation ${d\rho^2\over 4\rho^2}=dt^2$ so that one has
that $\rho=e^{-2t}$, $t\rightarrow \infty$ at  future infinity
${\cal I}^+$. The coordinate $t$ is convenient for measuring the
geodesic distance (the proper time) along a timelike geodesic. The
other coordinate is $\eta=e^{-t}$, $\eta \geq 0$, $\eta=0$ at
future infinity \footnote{Note that we are using  a convention in
which $\eta$  is positive and {\sl decreases} towards future
timelike infinity ${\cal I}^+$.}. In terms of the coordinate
$\eta$ the metric takes the form \be &&ds^2={1\over
\eta^2}\left(-d\eta^2+g_{ij}(x,\eta)dx^idx^j\right)\nonumber \\
&&g(x,\eta)=g^{(0)}(x)+g^{(2)}(x)\eta^2+g^{(3)}(x)\eta^3+.. \lb{6}
\ee where one has, as was first shown by Starobinsky
\cite{Starobinsky}, that \be &&g^{(2)}_{ij}(x)=R_{ij}-{1\over
4}Rg^{(0)}_{ij}~, ~~\Tr g^{(3)}=0~,~~\nabla^jg^{(3)}_{ij}=0~~,
\lb{7}\ee where the trace and covariant derivative are defined
with respect to metric $g^{(0)}_{ij}(x)$. Thus, starting with
$\eta^3$ there appear both even and odd powers of $\eta$.

\bigskip

\noindent{\it The Riemann coordinates.} Our coordinate system $\{
x^i \}$ on ${\cal I}^+$ should be adopted to a concrete observer
that follows a geodesic $\gamma$ parameterized by coordinate $t$.
Suppose that $\gamma$ intersect ${\cal I}^+$ at a point ${\cal O}$
with coordinates $x^i=0,\ i=1,..,d$. In a small vicinity of this
point one can choose the Riemann coordinate system (for a nice
introduction to this coordinate system see \cite{Petrov}) such
that \be &&g^{(0)}_{ij}(x)=\delta_{ij}-{1\over
3}R_{ikjn}(0)x^kx^n-{1\over 6}\nabla_k R_{injl}(0)x^k x^n
x^l+\dots
\nonumber \\
&&g^{(2)}_{ij}(x)=(R_{ij}(0)-{1\over
4}R(0)\delta_{ij})+\nabla_k(R_{ij}-{1\over
4}R\delta_{ij})(0)x^k+\dots \lb{8} \ee In terms of the spherical
coordinates $(r,\theta,\phi)$ with centre  at $x=0$, one has
$x^k=r n^k(\theta,\phi)$, $k=1,2,3$, where $n^k$ is unit vector,
$n^kn^k=1$.

\bigskip

\noindent{\it The causal diamond.} We choose the point $q$ to have
coordinates $(\eta=\epsilon,0,0,0)$ and point $p$ to have
coordinates $(\eta=N+\epsilon,0,0,0)$. In terms of coordinate $t$
we have that $t_\epsilon=\ln{1\over \epsilon}$ and
$t_{N+\epsilon}=\ln{1\over N+\epsilon}$ so that the proper time
interval is $\tau=t_\epsilon-t_{N+\epsilon}=\ln ({N+\epsilon\over
\epsilon})$. Notice that $\tau$ can be any finite number. In terms
of $\tau$ one has that $N=\epsilon (e^\tau-1)$. To leading order
the equation for the light-cone ${\dot I}^-(q) $ is
$$
r=\eta-\epsilon~~,\ 0\leq r\leq {N\over 2}
$$
while the equation for the light-cone ${\dot I} ^+(p) $ is
$$
r=N+\epsilon-\eta~~, \ 0\leq r\leq {N\over 2}~~.
$$

In our calculation we will need the next to leading order
modification of the light-cone. In metric (\ref{6}) the
null-geodesic satisfies equation \be {d\eta\over d\lambda}=\pm
\sqrt{g_{ij}(x,\eta)n^in^j}\ {dr\over d\lambda}\, , \lb{s1} \ee
where $\lambda$ is an affine parameter along the geodesic. To
second order in $r$ and $\eta$ one has that \be
g_{ij}(x,\eta)=\delta_{ij}-r^2{1\over
3}R_{ikjl}n^kn^l+\eta^2(R_{ij}-{1\over 4}R\delta_{ij}) \lb{9} \ee
so that
$$
g_{ij}n^in^j=1+\eta^2 (R_{ij}n^in^j-{1\over 4}R)\, .
$$
Substituting this into equation (\ref{s1}) and integrating we
find the equation for ${\dot I}^-(q) $, the past
 light-cone of $q$,   up to cubic order
in $\eta$, \be r=r_+(\epsilon)\equiv(\eta-\epsilon)-{1\over
6}(R_{ij}n^in^j-{1\over 4}R)(\eta^3-\epsilon^3)\, , \lb{s2} \ee
where we took into account the condition that $r=0$ when
$\eta=\epsilon$. A similar equation holds for ${\dot  I}^+(p) $, \be
r=r_-(\epsilon)\equiv (N+\epsilon-\eta)+{1\over
6}(R_{ij}n^in^j-{1\over 4}R)(\eta^3-(N+\epsilon)^3)\, . \lb{s22}
\ee The intersection of the two light-cones, ${\dot { I}^+(p)} \cap
{\dot {
I}^-(q)} $, is given by equation \be \eta={N\over 2}+\epsilon
-{1\over 8}(R_{ij}n^in^j-{1\over 4}R) N^2({N\over 2}+\epsilon)\, .
\lb{s23} \ee The correction to the flat space-time result is of
order $\epsilon^3$ and will be neglected in the calculation below.

\bigskip

\noindent{\it The volume.} Consider first the volume of the causal
diamond not taking into account the modification of the
light-cone. The volume inside the causal diamond is given by
expression \be V_1=\int_\epsilon^{{N\over 2}+\epsilon}{d\eta\over
\eta^4}\int_0^{\eta-\epsilon}dr ~r^2\int_{S_2}\sqrt{\det~ g}+
\int^{N+\epsilon}_{{N\over 2}+\epsilon}{d\eta\over
\eta^4}\int_0^{N+\epsilon-\eta}dr ~r^2\int_{S_2}\sqrt{\det~ g}
\lb{10}\ee where we introduced
$$
\int_{S_2}=\int_0^\pi d\theta\sin\theta \int_0^{2\pi}d\phi~~.
$$
To the second order one has that \be \sqrt{\det g}=1-{r^2\over
6}R_{kl}(0)n^k(\theta,\phi)n^l(\theta,\phi)+{\eta^2\over
8}R(0)+\dots \lb{11} \ee One checks by direct calculation that \be
\int_{S_2}n^kn^l={4\over 3}\pi \delta^{kl} \, .\lb{12}\ee One thus
finds for the volume \be V_1={4\pi\over 3}J_1(\tau)-{2\pi\over
45}R(0)\epsilon^2J_2(\tau)+{\pi\over 6}R(0)\epsilon^2J_3(\tau)\, .
\lb{13} \ee

The contribution to the volume due to the modifications (\ref{s2})
and (\ref{s22})  of  light-cones ${\dot I}^+(p)$ and ${\dot
I}^-(q)$ is given by expression \be
V_2=\int_{S_2}\left(\int_\epsilon^{{N\over 2}+\epsilon}{d\eta\over
\eta^4}\, \int_{\eta-\epsilon}^{r_{+}(\epsilon)} dr \,
r^2+\int_{{N\over 2}+ \epsilon}^{N+\epsilon}{d\eta\over \eta^4}\,
\int_{\eta-\epsilon}^{r_{-}(\epsilon)} dr \, r^2\right)\, .
\lb{v2} \ee Keeping only  terms quadratic in $\epsilon$ and using
identity (\ref{12}) we find that \be V_2=-{\pi\over
18}R(0)\epsilon^2J_4(\tau)\, , \lb{s3} \ee where we introduced
(recall that $N=\epsilon (e^\tau-1)$) \be &&\int_\epsilon^{{N\over
2}+\epsilon}d\eta ({1\over
\eta^4}+{1\over (N+2\epsilon-\eta)^4})(\eta-\epsilon)^3=J_1(\tau)\lb{14} \\
&&\int_\epsilon^{{N\over 2}+\epsilon}d\eta({1\over
\eta^4}+{1\over (N+2\epsilon-\eta)^4})(\eta-\epsilon)^5=\epsilon^2J_2(\tau)\nonumber \\
&&\int_\epsilon^{{N\over 2}+\epsilon}d\eta ({1\over
\eta^2}+{1\over (N+2\epsilon-\eta)^2})(\eta-\epsilon)^3=\epsilon^2J_3(\tau) \nonumber \\
&&\int_\epsilon^{{N\over 2}+\epsilon}d\eta
({(\eta^3-\epsilon^3)\over
\eta^4}-{(N+2\epsilon-\eta)^3-(N+\epsilon)^3\over
(N+2\epsilon-\eta)^4})(\eta-\epsilon)^2=\epsilon^2 J_4(\tau)\, .
\nonumber \ee The integration can be performed explicitly so that
one gets the closed form expressions \be
J_1(\tau)&=&2\ln\cosh{\tau\over
2}-\tanh^2({\tau\over 2}) \nonumber \\
J_2(\tau)&=&{5\over 12}(17e^{2\tau}+38e^\tau+17)\tanh^2({\tau\over
2})+10(e^{2\tau}+1)\ln ({1+e^{-\tau}\over 2})+10\tau \nonumber \\
J_3(\tau)&=&{9\over 4}(e^\tau-1)^2+3(e^{2\tau}+1)\ln ({1+e^{-\tau}\over 2})+3\tau \nonumber \\
J_4(\tau)&=&{1\over
12}(13e^{2\tau}+10e^{\tau}+13)\tanh^2({\tau\over
2})+(e^{2\tau}+1)\ln ({1+e^{-\tau}\over 2})+\tau \lb{JJ} \ee
Notice that there is an  identity which holds for these functions \be -{4\over
5}J_2(\tau)+3J_3(\tau)-J_4(\tau)=0\, . \lb{id} \ee

Recalling that $\epsilon=e^{-t_q}$ where $t_q$ is the time
coordinate of the point $q$ and combining all contributions we
obtain the volume as expansion in powers of $e^{-t_q}$, \be
V&=&V_1+V_2=
a_0(\tau)+a_2(\tau)e^{-2t_q}+a_4(\tau) e^{-4t_q}+\dots ~~,\nonumber \\
a_0(\tau)&=&{4\pi\over 3}J_1(\tau)\, , \  a_2(\tau)=(-{2\pi\over
45}J_2(\tau)+{\pi\over 6}J_3(\tau)-{\pi \over 18}J_4(\tau))R(0)\,
, \lb{16} \ee where $R(0)$ is the Ricci scalar of the
3-dimensional surface ${\cal I}^+$ at the point of intersection of
the geodesic $\gamma$ with ${\cal I}^+$. Notice that expansion in
$e^{-2t_q}$ is also expansion in the curvature (and its
derivatives) of ${\cal I}^+$. Now, it is a surprising fact that
due to identity (\ref{id})  the coefficient $a_2(\tau)$ vanishes
identically, \be a_2(\tau)\equiv 0\, . \lb{a2} \ee

Notice that the possible term in (\ref{16}) which is  cubic in
$e^{-t_q}$ vanishes. This is due to the fact that in the expansion
(\ref{6}) one has that $\Tr g^{(3)}=0$ and due to  the property
$$
\int_{S_2}n^k n^l n^m=0~~.
$$
The other possible source for a $e^{-3t_q}$ term is the
$\epsilon^3$ modification in (\ref{s23}). The analysis however
shows that this modification shows up in the volume in the form of
even powers of $\epsilon^3$, i.e. it may first appear in term
$e^{-6t_q}$.

So that the next non-vanishing term in expansion (\ref{16}) is
$e^{-4t_q}$. It would be interesting to see whether all odd powers
of $e^{-t_q}$ vanish in expansion (\ref{16}) of the volume.
 We  note that the volume has finite limit when $t_q\rightarrow
\infty$ so no regularization is needed.  At first sight, this is
surprising since taking that the volume of a bulk region is
typically divergent when the boundary of the region approaches
infinity (spacelike in anti-de-Sitter and timelike in de-Sitter
space-time, see \cite{Henningson:1998gx} and
\cite{deHaroSolodukhinSkenderis}). However in maximally  symmetric
spacetime like de-Sitter, it is clear that all causal diamonds
with the same duration $\tau$ are equivalent, no matter how close
to future infinity ${\cal I} ^+$ they may be,   and they  must
therefore have the same volume $V(\tau)$ given in fact by the
universal formula for  $J_1(\tau)$ in  (\ref{JJ}) . The same
universal formula was obtained in a different way in
\cite{GibbonsSolodukhin} (see (21)  of that reference).

We emphasize that the first term in the  expansion (\ref{16})
comes from the metric \be
ds^2=-dt^2+e^{2t}(dr^2+r^2(d\theta^2+\sin^2\theta)) \lb{17} \ee of
de-Sitter spacetime with flat constant $t$ slices. This is the
only contribution in the limit of $t_q\rightarrow \infty$. The
curvature of the spacelike surface ${\cal I}^+$ shows up in the
$e^{-2nt_q}$ correction terms. Thus the information on the
curvature of ${\cal I}^+$ which is encoded in the volume of the
causal diamond is exponentially suppressed. When the diamond as a
whole moves closer to the future infinity the geometry inside the
diamond becomes more and more accurately de-Sitter. This is of
course consistent with results of \cite{BoucherGibbons}. There is
a nice universality: no matter what is the local geometry in the
bulk the geometry inside the diamond becomes de-Sitter when it
approaches the future infinity.

At fixed duration $\tau$ the volume of causal diamond in pure de
Sitter space-time becomes a function of the cosmological constant
$\Lambda$, \be V_{\tt dS}(\tau,\Lambda)={4\pi\over
\Lambda^2}(2\ln\cosh({\tau\sqrt{3\Lambda}\over
6})-\tanh^2({\tau\sqrt{3\Lambda}\over 6}))\, . \ee
\begin{figure}
\centerline{\rotatebox{270}{\epsfig{figure=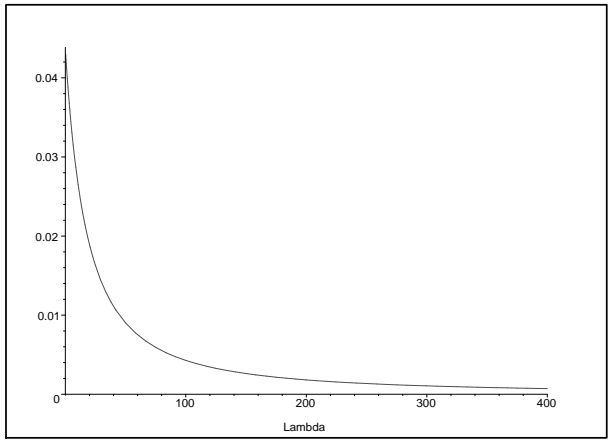, height=8cm}}}
\caption{The volume of causal diamond of duration $\tau=1$ in pure
de-Sitter space-time as function of cosmological constant
$\Lambda$. }
\end{figure}
In models of eternal inflation $V(\tau, \Lambda)$ is taken as a
measure of probability of an observer of duration $\tau$. Thus, we
can see how this probability depends on cosmological constant
$\Lambda$. As is seen in Figure 1 the volume is monotonically
decreasing with $\Lambda$ taking the maximal value at vanishing
$\Lambda$.

\section{Relation to the quasi-normal modes}

\begin{figure}
\centerline{\rotatebox{270}{\epsfig{figure=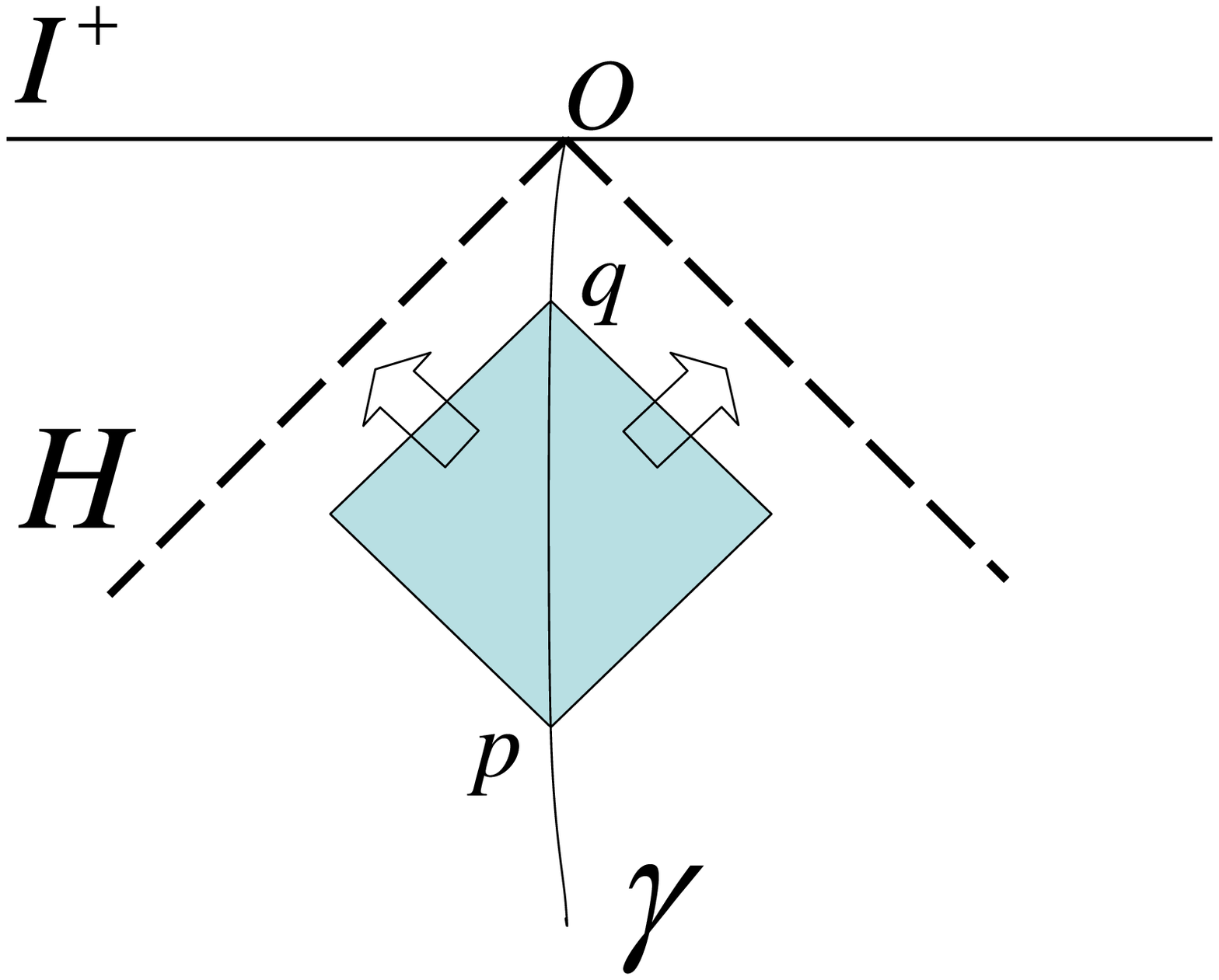,
height=8cm}}} \caption{The causal diamond when approaching the
future infinity ${\cal I}^+$ becomes more and more accurately
described by static de-Sitter coordinates inside the horizon
${\cal H}$ associated with the observer following the timelike
geodesic $\gamma$. The gravitational perturbations over the de
Sitter metric is radiated through the boundary of the diamond. The
 size of the diamond on the conformal diagram becomes smaller and
 smaller
when $q$ and $p$ approach $\cal O$. }
\end{figure}

There is an alternative way of looking at the time evolution of
the geometry inside the causal diamond. In the limit $t_q,\, t_p
\rightarrow \infty$ the diamond is close to the corner formed by
cosmological event horizon ${\cal H}$ of the observer that follows
the timelike geodesic $\gamma$. Inside this corner one can always
take  the de-Sitter metric in static coordinate system  as a
background and consider deviations from the de-Sitter space as
perturbations. A perturbation is described by a wave equation and
takes  the form ${1\over r} H_l(r) e^{-i\omega t_S} Y_l(\theta,
\phi)$, where $H_l(r)$ satisfies an effective radial Shr\"odinger
type equation. Inside the diamond these perturbations tend to
escape through the boundaries of the diamond. In a bigger picture
the perturbations dissipate through the event horizon ${\cal H}$.
The concrete mechanism of the dissipation is given by the
quasi-normal modes which are solutions to the gravitational
equations for the perturbations subject to condition that they are
out-going at the horizon and regular at the origin. This condition
can be satisfied only for a discrete complex set of frequencies
$\omega_n$. For de-Sitter space-time the gravitational
quasi-normal modes have been studied for instance in
\cite{Brady:1999wd} and \cite{dS}. An interesting peculiarity of
de-Sitter spacetime as compared to a black hole spacetime is that
the quasi-normal frequencies are purely imaginary so that they
describe the exponential decay only while generically there could
be also oscillations\footnote{The discussion in the literature of
the quasi-normal modes  in pure de-Sitter spacetime is
controversial. This is because there is an apparent cancellation
of the would-be quasi-normal poles if $i\omega$ is an integer and
spacetime is pure de-Sitter. On the other hand, as was noted in
\cite{Brady:1999wd} the presence of an arbitrary small (but
non-vanishing) black hole mass prevents $i\omega$ from being an
integer and leads to a negligible correction to the quasi-normal
modes.}. In D spacetime dimensions there are two sets of the
quasi-normal modes \cite{dS} \be \omega_n=-i(l+D-1-q+2n)\, ,
~~\omega_n=-i(l+q+2n)\, , \ \lb{20} \ee where $n=0,1,2,..$; \ $l$
is the angular momentum of the perturbation and the value of $q$
depends on the type of the perturbation: $q=0$ for tensor, $q=1$
for vector and $q=2$ for scalar perturbations.

The perturbation of the volume of the causal diamond \be \delta
V=\int_{\diamondsuit}{1\over 2}\sqrt{G}G^{\mu\nu}H_{\mu\nu}
\lb{21} \ee is determined by a scalar type gravitational
perturbation.
Moreover, since the integration in (\ref{21}) includes integration
over spherical angles then only $l=0$ may contribute to the time
evolution of the volume.

Let us now compare the two sets (for $D=4$) of frequencies
(\ref{20}) with our direct calculation (\ref{16}). Doing this one
should keep in mind the relation between global and static
coordinate systems. One has that $\sinh t=(1-r^2)\sinh {t_S}$,
where $t_S$ is time coordinate in static coordinate system and $t$
is time coordinate in the global coordinate system. Inside the
diamond $r$ changes in the finite limits. Therefore, for large
times $t\sim t_S$.
 We see that
the set in (\ref{20}) in which $i\omega$ is an odd number does not
show up in the evolution of the volume. At least this is true for
the few lowest  frequencies. On the other hand, the second set, in
which $i\omega$ is even number, indeed appears in the evolution of
the volume.  We  have repeated the calculation in the previous
section for arbitrary $D$. Then the lowest decaying (and,
possibly, non-vanishing) term in the volume (\ref{16}) is   of
order $e^{-2t_q}$. This is again consistent with the second set of
quasi-normal frequencies in (\ref{20}).

\section{Acknowledgements}
Ths work was initiated at  the IH\'ES. Both authors would like to
thank Thibault Damour and the director Jean Pierre Bourguignon,
for their hospitality during our stay at IH\'ES. The work was
completed while the first author was visiting the Galileo Gallilei
Institute and he would like to thank the director and the
organisers of the the workshop on \lq String and M theory
approaches to particle physics and cosmology\rq{}  for their
hospitality and INFN for partial support. The second author would
like to thank Michael Gromov for an interesting discussion.

\end{document}